\documentclass[]{article}

\usepackage{moreverb}
\usepackage[colorlinks=true,bookmarksopen,bookmarksnumbered,linkcolor=black,citecolor=black,urlcolor=black]{hyperref}
\usepackage[ngerman, UKenglish]{babel}
\usepackage{amsmath}
\usepackage{amsthm}
\usepackage{amsfonts}
\usepackage{placeins}
\usepackage[usenames,dvipsnames]{xcolor}
\usepackage{tikz}
\usepackage[european]{circuitikz}
\usepackage[acronym]{glossaries}
\usepackage{graphicx}
\usepackage{caption}
\usepackage{subcaption}
\usepackage{float}
\usepackage{tabularx}
\usepackage{multirow}
\usepackage{bigstrut}
\usepackage{authblk}

\makeglossaries
\DeclareMathAlphabet\mathbfcal{OMS}{cmsy}{b}{n}

\newcommand\BibTeX{{\rmfamily B\kern-.05em \textsc{i\kern-.025em b}\kern-.08em
        T\kern-.1667em\lower.7ex\hbox{E}\kern-.125emX}}

\newacronym{ic}{IC}{integrated circuit}
\newacronym{em}{EM}{electromagnetic}
\newacronym{fit}{FIT}{finite integration technique}
\newacronym{mc}{MC}{Monte Carlo}
\newacronym{qoi}{QoI}{quantities of interest}
\newacronym{pdf}{PDF}{probability density function}
\newacronym{rv}{RV}{random variable}
\newacronym{stc}{SC}{stochastic collocation}
\newacronym{sg}{SG}{sparse grid}
\newacronym{svd}{SVD}{singular value decomposition}
\newacronym{tt}{TT}{tensor train}
\newacronym{uq}{UQ}{uncertainty quantification}

\title{High Dimensional Uncertainty Quantification for an Electrothermal Field Problem using Stochastic Collocation on Sparse Grids and Tensor Train Decompositions}

\author[1,2]{D.~Loukrezis}
\author[1,2]{U.~R\"omer}
\author[1,2]{T.~Casper}
\author[1,2]{S.~Sch\"ops}
\author[1,2]{H.~De~Gersem}
\affil[1]{ \small Technische Universit\"at Darmstadt, Institut f\"ur Theorie Elektromagnetischer Felder, Schlo\ss gartenstra\ss e 8, D-64289, Darmstadt, Germany}
\affil[2]{Technische Universit\"at Darmstadt, Graduate School for Computational Engineering, Dolivostra\ss e 15, D-64293, Darmstadt, Germany}

\date{\vspace{-5ex}}

\begin{document}

\maketitle

\providecommand{\keywords}[1]{\textbf{\textit{keywords---}} #1}

\begin{abstract}
The temperature developed in bondwires of \glspl{ic} is a possible source of malfunction, and has to be taken into account during the design phase of an \gls{ic}.
Due to manufacturing tolerances, a bondwire's geometrical characteristics are uncertain parameters, and as such their impact has to be examined with the use of \gls{uq} methods.
Considering a stochastic electrothermal problem featuring twelve (12) bondwire-related uncertainties, we want to quantify the impact of the uncertain inputs onto the temperature developed during the duty cycle of an \gls{ic}. 
For this reason we apply the \gls{stc} method on \glspl{sg}, which is considered the current state-of-the-art.
We also implement an approach based on the recently introduced low-rank tensor decompositions, in particular the \gls{tt} decomposition, which in theory promises to break the curse of dimensionality.
A comparison of both methods is presented, with respect to accuracy and computational effort.

\keywords{electrothermal field simulation, \gls{ic} packaging, low-rank tensor decompositions, sparse grids, stochastic collocation, tensor trains, uncertainty quantification.}
\end{abstract}

\section{Introduction}

A thorough electrothermal analysis is a key step in the design of integrated circuits (\glspl{ic}). 
One important aspect is the Joule heating of the bondwires which connect the \gls{ic} to its surrounding packaging, as wire fusing is a common source of failure. 
Thermal heating is affected by uncertainties introduced in the manufacturing process, e.g. with respect to the bondwire's length. 
More often than ever before \glsfirst{uq} methods are applied, with the aim of assessing the risk of fusing or failure. 
In this way, one may avoid costly and time-consuming over-engineering.

In a stochastic setting, state-of-the-art UQ methods are the stochastic Galerkin and collocation method \cite{BabSG, XiuHigh}. 
These methods use a spectral approximation based on a global multivariate polynomial basis. Given a high regularity of the solution with respect to the random input variables, fast convergence rates can be achieved. 
Hence, the computational cost of using stochastic Galerkin and collocation can be affordable even for applications with a high complexity. 

This might no longer be true as the number of random inputs increases. 
Referring to the bondwire case, the number of wires can be large and, at the same time, randomness hampers the exploitation of any symmetry properties, which could otherwise be used in order to reduce the computational effort.
The difficulty arising when approximating high-dimensional problems is called the curse of dimensionality \cite{Bellmann}. 
In the case of standard Galerkin and collocation methods, this cannot simply be overcome with a further increase of computational power, as the complexity scales exponentially with the number of random inputs. 
Hence, the construction of more dedicated schemes is in order. 

The efficiency of stochastic spectral methods can be enhanced for high-dimensional problems using sparse approximations. 
In this case, dimension independent approximation rates can be derived under very strict assumptions on the solution regularity and the input randomness \cite{ChkBreak}. 
Non-intrusive collocation based on sparse grids \cite{BarSG, BungSparse, NobSparse} outperforms Monte Carlo and polynomial tensor approximations when the underlying functions possess high (mixed) regularity. 
However, if some of the assumptions are not fully satisfied, the convergence of sparse approximations is generally not free of dimensionality problems. 
A recent application of sparse approximations in electromagnetics is described in \cite{NguLeast} based on the least angle regression method. 

Recently, based on the observation that multivariate polynomial bases have an underlying tensor structure, low-rank tensor approximations have gained increasing interest \cite{GrasRev, HackBook, KhorRev, KoldaRev}.
First results on low-rank tensor approximations show an improved asymptotic convergence rate with respect to the number of random inputs  compared to sparse grids \cite{UscPhD}, yet many theoretical questions persist. 
Moreover, the computational effort for more complicated examples has hardly been investigated.
It is the aim of the present paper to contribute in this direction. 

This paper compares the accuracy and computational complexity of stochastic collocation on sparse grids and the one of tensor train decomposition. 
Both methods are considered in the context of electrothermal field simulations, e.g. occurring in \gls{ic} packaging problems, characterized by high dimensional uncertainty, i.e. with a comparably large number of uncertain parameters.

The rest of the paper is structured as follows: 
Section 2 deals with an electrothermal field problem and its numerical discretization. 
In Section 3 the considered uncertainties and quantities of interest are described. 
Section 4 describes both stochastic collocation on sparse grids and low-rank tensor approximations using tensor trains. 
A comparison of both methods when applied to the considered electrothermal problem is presented in Section 5.

\section{Electro-thermal Field Problem} \label{sec:ETProblem}
Due to the Joule heating effect, the temperature of a wire increases if an electrical current is applied to it. 
In turn, temperature changes affect the wire's material parameters. 
Even if the dependence of the volumetric heat capacity on the temperature is neglected, the nonlinearity in electrical and thermal conductivities with respect to temperature remains.
This electrothermal coupling effect is presented in the following subsections, for the continuous and the discrete case, respectively.

\subsection{Continuous Problem}
The electrical part of the problem can be described with the current continuity equation.
If capacitive effects are ignored, it suffices to write the electro-kinetic problem with suitable boundary conditions
\begin{align}
\label{eq:electric}
\nabla \cdot \sigma(\cdot, T) \nabla\phi &= 0  &&\text{in} \:\:\:\: D \times I, \nonumber \\
\phi_i &= V_i  &&\text{on} \:\:\:\: \Gamma_{\text{Dir}, i} \times I,  \\
\sigma(\cdot, T) \nabla\phi \cdot \vec{n} &= 0  &&\text{on} \:\:\:\: \left(\Gamma \setminus \Gamma_{\text{Dir}}\right) \times I, \nonumber
\end{align}
where $D$ denotes the computational domain, $I$ the time interval, $\Gamma$ the boundary of the computational domain, $\Gamma_{\text{Dir}}$ the Dirichlet boundary, $V_i$ the potential applied on the $i$-th contact pad, $\vec{n}$ the outer normal vector, $\sigma$ the electrical conductivity, $\phi$ the electric potential and $T$ the temperature.

Thermal heat flow can be attributed to conduction, convection and radiation.
The thermodynamic power balance is described with the transient heat equation, which reads
\begin{align}
\label{eq:thermal}
\rho c \: \partial_t T  - \nabla \cdot \lambda(\cdot, T) \nabla T &= q(\cdot, T, \phi) &&\text{in} \:\:\:\: D \times I, \nonumber \\
\lambda(\cdot, T) \nabla T \cdot \vec{n} &= \ h T + \varepsilon \sigma_{\text{SB}} T^4  - h T_{\infty} - \varepsilon \sigma_{\text{SB}} T_{\infty}^4 &&\text{on} \:\:\:\: \Gamma \times I, \\
T &= T_{\infty}, &&\text{on} \:\:\:\: \Gamma \times \{0\}, \nonumber
\end{align}
where $\rho c$ denotes the volumetric heat capacity, $\lambda$ the thermal conductivity, $q$ the power density of the system's heat sources, $h$ the heat transfer coefficient, $\varepsilon$ the emissivity, $\sigma_{\text{SB}}$ the Stefan-Boltzmann constant and $T_{\infty}$ the ambient temperature. 
The heat sources comprise of the electrical contribution $q_{\text{el}}$ due to Joule heating and the bondwire contribution $q_{\text{bw}}$, such that $q = q_{\text{el}} + q_{\text{bw}}$.
Heat exchange with the environment is modeled as convective and radiative boundary conditions, see \cite{CasEle}.

The combination of the transient heat equation with the electro-kinetic problem yields a nonlinear electrothermal system, where the electrothermal coupling is established due to the dependencies of the material parameters $\lambda$ and $\sigma$ on the temperature $T$ and of the heat source $q$ on the electric potential $\phi$.
The Joule heating due to the electro-kinetic problem reads
\begin{equation}
q_{\text{el}} = (\nabla \phi)^{\top} \sigma(T) \nabla \phi.
\end{equation}

\subsection{Discrete Problem} \label{subsec:Discrete}
The coupled electrothermal problem is discretized in space with the \gls{fit} \cite{WeiTime} on a primal and a dual Cartesian mesh.
The discrete unknowns, i.e. the electric potentials $\mathbf{\Phi}$ and the temperatures $\mathbf{T}$, are allocated at the nodes of the primary grid.
Omitting specific implementation details and initial conditions, the discrete analogue to (\ref{eq:electric}), (\ref{eq:thermal}) reads 
\begin{align} 
\label{disc1}
-\tilde{\mathbf{S}} \mathbf{M}_{\sigma}(\mathbf{T}) \mathbf{G} \mathbf{\Phi} &= \mathbf{0},  \\
\label{disc2}
\mathbf{M}_{\rho c} \mathbf{\dot{T}} - \tilde{\mathbf{S}} \mathbf{M_{\lambda}}(\mathbf{T}) \mathbf{G} \mathbf{T} &= \mathbf{Q}(\mathbf{T}, \mathbf{\Phi}),
\end{align}
where $\tilde{\mathbf{S}}^{\top} = - \mathbf{G}$ is a topological operator, with $\mathbf{G} \in \{-1, 0, 1\}^{n \times n}$ being the edge-node incidence matrix, $\mathbf{M}_{\sigma}$ the electrical conductance mass matrix, $\mathbf{M}_{\lambda}$ the thermal conductance mass matrix, $\mathbf{M}_{\rho c}$ the thermal capacitance mass matrix and $\mathbf{Q}$ the discrete counterpart of the heat sources.

\begin{figure}[b!]
    \centering
    \begin{tikzpicture}[xscale=1.0]
    \draw (0,0) -- +(30:2cm);
    \draw (1,0) -- +(30:2cm);
    \draw (2,0) -- +(30:2cm);
    \draw (3,0) -- +(30:2cm);
    \draw (0,0) -- (3,0);
    \draw (1.732,1) -- (4.732,1);
    \draw (0.866,0.5) -- (3.886,0.5);
    \draw[line width=2pt, line cap=round, color=blue] (0.866,0.5) .. controls(2.5,2.5) and (2.0,0.8) .. (3.886,0.5);
    \path (4.9,0.5) node{\huge $\rightarrow$};  
    \draw (5,0) -- +(30:2cm);
    \draw (6,0) -- +(30:2cm);
    \draw (7,0) -- +(30:2cm);
    \draw (8,0) -- +(30:2cm);
    \draw (5,0) -- (8,0);
    \draw (6.732,1) -- (9.732,1);
    \draw (5.866,0.5) -- (8.886,0.5);
    \draw (5.866,0.5) -- (5.866,1.7) to[generic, l=$G_{\text{bw}}^{\text{el}}(T)$, color=blue] (8.886,1.7) -- (8.886,0.5);
    \end{tikzpicture}
    \caption{Bondwire lumped element modeling of the electrical conductance.}
    \label{fig:lumped}
\end{figure}
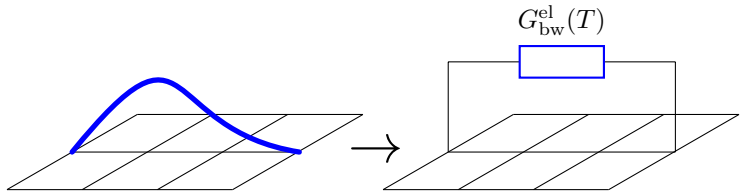

Bondwires are not resolved by the \gls{fit} grid, but are modeled with a lumped element approach, where pairs of mesh points are connected by an electrothermal lumped element.
This approach is illustrated in Figure \ref{fig:lumped}.
Both conductances depend on the wire's material properties and geometrical dimensions, i.e.,
\begin{align}
G_{\text{bw},j}^{\text{el}} = \frac{\sigma_{\text{bw},j} A_{\text{bw},j}}{\ell_{\text{bw},j}} 
\:\:\:\:\:\:\:\:\:\:\:\: \text{and} \:\:\:\:\:\:\:\:\:\:\:\:
G_{\text{bw},j}^{\text{th}} = \frac{\lambda_{\text{bw},j} A_{\text{bw},j}}{\ell_{\text{bw},j}},
\label{Gbw}
\end{align}
where $\ell_{\text{bw},j}$ denotes the length of the $j$-th bondwire and $A_{\text{bw}, j}$ its cross-section surface.
The conductance terms $G_{\text{bw},j}^{\text{el}}$, $G_{\text{bw},j}^{\text{th}}$ have to be assembled into the system matrices of (\ref{disc1}), (\ref{disc2}), based on the bondwires' connection positions.
Then, the extended electrothermal system reads
\begin{align}
\label{eq:ext1}
-\tilde{\mathbf{S}} \mathbf{M}_{\sigma}(\mathbf{T}) \mathbf{G} \mathbf{\Phi} &+ \sum_{j=1}^{N_{\text{bw}}} \mathbf{P}_j G_{\text{bw}, j}^{\text{el}}(T_{\text{bw}, j}) \mathbf{P}_j^{\top} \mathbf{\Phi} = \mathbf{0}, \\ 
\label{eq:ext2}
\mathbf{M}_{\rho c} \mathbf{\dot{T}} - \tilde{\mathbf{S}} \mathbf{M}_{\lambda}(\mathbf{T}) \mathbf{G} \mathbf{T} &+ \sum_{j=1}^{N_{\text{bw}}} \mathbf{P}_j G_{\text{bw}, j}^{\text{th}}(T_{\text{bw}, j}) \mathbf{P}_j^{\top} \mathbf{T} = \mathbf{Q}(\mathbf{T}, \mathbf{\Phi}),
\end{align}
where $N_{\text{bw}}$ denotes the amount of bondwires, $\mathbf{P}_j \in \{-1, 0, 1\}^{n \times n}$ is a bondwire gradient vector which handles the incidences between the contacts of the $j$-th bondwire and the grid, and $T_{\text{bw}, j} = \mathbf{X}_j^{\top} \mathbf{T}$ is the average temperature of the $j$-th bondwire, with vector $\mathbf{X}_j$ being used to average the temperature from both connection nodes of the bondwire.
The bondwires' contribution to the heat sources reads
\begin{align}
\label{eq:Q_bw}
\mathbf{Q}_{\text{bw}} &= \sum_{j=1}^{N_{\text{bw}}} \mathbf{X}_j Q_{\text{bw}, j}, \\
\text{where} \:\:\: Q_{\text{bw}, j} &= \boldsymbol{\Phi}^{\top} \mathbf{P}_j G_{\text{bw}, j}^{\text{el}}(T_{\text{bw}, j}) \mathbf{P}_j^{\top} \boldsymbol{\Phi}. \nonumber
\end{align}

\section{Uncertainties and Quantities of Interest} 
% input uncertainties part
\subsection{Input Uncertainties}
The bondwire's temperature depends on the wire's material properties, its geometry and on the applied electric voltage. 
Assuming that the electric voltage can be accurately controlled and that material properties are known with high precision, the only remaining source of uncertainty is related to the wire's geometry. 
Further assuming that a wire can be fabricated with an almost exact cross-section, the wire's length remains as the only uncertain parameter.
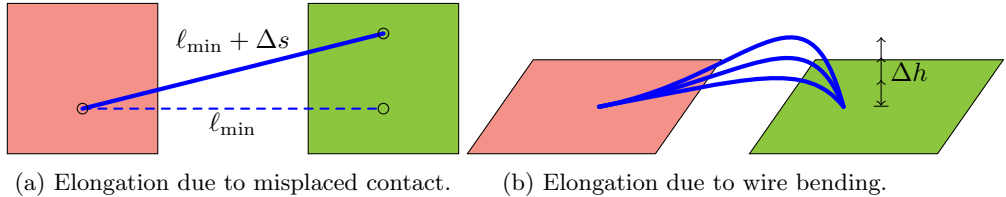
\begin{figure}[b!]
    \begin{subfigure}[b]{0.5\textwidth}
        \centering
        \tikzset{cross/.style={cross.out, draw=black, minimum size=2*(#1-\pgflinewidth), inner sep=0pt, outer sep=0pt},
            %default radius will be 1pt. 
            cross/.default={1pt}}
        \begin{tikzpicture}[scale=2.0]
        % draw contact pad, chip and wire
        \draw[fill=Salmon] (0,0) rectangle (1,1);
        \draw[fill=LimeGreen] (2,0) rectangle (3,1);
        \draw[thick,color=blue,line cap=round, dashed] (0.5, 0.3) -- (2.5, 0.3);
        \draw[ultra thick,color=blue,line cap=round] (0.5, 0.3) -- (2.5, 0.8);
        % draw crosses
        \draw (0.5,0.3) circle (1pt);
        \draw (2.5,0.3) circle (1pt);
        %\draw (2.5,0.8) circle (1pt);
        \draw (2.5,0.8) circle (1pt); %node[cross=3pt] {};
        % place text
        \node at (1.5,0.2) {$\ell_{\text{min}}$};
        \node at (1.5,0.75) {$\ell_{\text{min}} + \Delta s$};
        \end{tikzpicture}
        \caption{Elongation due to misplaced contact.}   
        \label{fig:subfigA}
    \end{subfigure}
    %%%
    \begin{subfigure}[b]{0.5\textwidth}
        \centering
        \begin{tikzpicture}[scale=2.5]
        \draw[fill=Salmon]
        (0.0, 0.5)		% starting point
        -- ++(0.35, 0.5)	% move along this vector
        -- ++(1, 0) 	        % then along this vector
        -- ++(-0.35, -0.5)	% then back along that vector
        -- cycle;		% and back to where you started
        \draw[fill=LimeGreen]
        (1.5, 0.5)
        -- ++(0.35,0.5)
        -- ++(1,0)
        -- ++(-0.35,-0.5)
        -- cycle;
        \draw[ultra thick,line cap = round,color=blue] (0.7,0.75) .. controls(1.5,0.9) and (1.8,1.5) .. (2,0.75);
        \draw[ultra thick,line cap = round,color=blue] (0.7,0.75) .. controls(1.5,0.9) and (1.8,1.25) .. (2,0.75);
        \draw[ultra thick,line cap = round,color=blue] (0.7,0.75) .. controls(1.5,0.9) and (1.8,1.0) .. (2,0.75);
        
        % draw measurement annotation
        \draw[|<->] (2.2,0.75) -- node[right]{$\Delta h$} (2.2,1.12);
        \draw[->] (2.2,0.75)  (2.2,1.0);
        \draw[->] (2.2,0.75)  (2.2,0.88);
        %\draw[->] (2.2,0.25) -- (2.2,0.8);
        %\draw[->] (2.2,0.25) -- (2.2,0.6);
        \end{tikzpicture}
        \caption{Elongation due to wire bending.}
        \label{fig:subfigB}
    \end{subfigure}
    \caption{Bondwire length variations due to fabrication tolerances.} 
    \label{fig:bw_uncertainty}
\end{figure}

A wire's length depends on the bonding process during the \gls{ic} fabrication.
The distance between the contact pad and the chip determines the minimum length of a wire, $\ell_{\text{min}}$.
Any deviation from the ideal position on the contact pad adds an elongation $\Delta s$ to the minimum length of the wire.
Any bending of the wire results in an additional elongation $\Delta h$, yielding the total wire length 
\begin{equation}
\ell_{\text{tot}} = \ell_{\text{min}} + \Delta s + \Delta h.
\end{equation}
These elongations are illustrated in Figure \ref{fig:bw_uncertainty}.
Due to the uncertainties concerning pad positions and wire bending, the wire's length can be considered as an uncertain parameter.
Instead of the total wire length, we choose the uncertain parameter to be the relative elongation, which is defined as 
\begin{equation}
\delta = \left( \ell_{\text{tot}} - \ell_{\text{min}} \right) / \ell_{\text{tot}}.
\end{equation} 
Specifically, each bondwire's relative elongation is drawn from the range $[0.122, 0.218]$ \cite{CasEle} in a uniform manner.
Hence, we consider one \gls{rv} $Y_n$ per bondwire in the \gls{ic}, corresponding to the random relative elongation and following a uniform distribution, i.e. $Y_n \sim \mathcal{U}[0.122, 0.218]$.
\glspl{rv} $Y_n$ are assumed to be stochastically \textit{independent}. 
Then the input \glspl{rv} are given as $\mathbf{Y} = (Y_1, Y_2, ..., Y_{N_{\text{bw}}})$, with joint \gls{pdf} $\varrho \left( \mathbf{y} \right)$.
Here, $\mathbf{y} = \mathbf{Y}(\omega)$ denotes a realization of the input \glspl{rv}, with $\omega \in \Omega$ being a random outcome, such that $\mathbf{Y} : \Omega \rightarrow \mathbb{R}^{N_{\text{bw}}}$.

% QoI part
\subsection{Output Uncertainties and Quantities of Interest}
Uncertainty in the inputs of the problem also leads to uncertainty in the discrete unknowns through the stochastic problem
\begin{align}
\label{eq:ext1_uq}
-\tilde{\mathbf{S}} \mathbf{M}_{\sigma}(\mathbf{T}) \mathbf{G} \mathbf{\Phi} &+ \sum_{j=1}^{N_{\text{bw}}} \mathbf{P}_j G_{\text{bw}, j}^{\text{el}}(\mathbf{Y},T_{\text{bw}, j}) \mathbf{P}_j^{\top} \mathbf{\Phi} = \mathbf{0}, \\ 
\label{eq:ext2_uq}
\mathbf{M}_{\rho c} \mathbf{\dot{T}} - \tilde{\mathbf{S}} \mathbf{M}_{\lambda}(\mathbf{T}) \mathbf{G} \mathbf{T} &+ \sum_{j=1}^{N_{\text{bw}}} \mathbf{P}_j G_{\text{bw}, j}^{\text{th}}(\mathbf{Y},T_{\text{bw}, j}) \mathbf{P}_j^{\top} \mathbf{T} = \mathbf{Q}(\mathbf{Y},\mathbf{T}, \mathbf{\Phi}).
\end{align}
In turn, the uncertainty in the solution has to be taken into consideration for the estimation of \gls{qoi}. 
Here, we are interested in possible sources of malfunction, i.e. output values which can lead to performance degradation or even to the failure of the \gls{ic}.
The maximum temperature $T_{\text{max}}$ developed in the \gls{ic}'s bondwires is the most probable cause for such malfunctions, hence  we choose $T_{\text{max}}$ to be the \gls{qoi}.
Temperature $T_{\text{max}}$ is a functional of the discrete solution of the electrothermal problem, i.e. $T_{\text{max}} = f(\mathbf{T})$.
A bondwire's temperature depends on the random relative elongation described above through \eqref{eq:ext1_uq} and \eqref{eq:ext2_uq}. Hence, denoting a \gls{qoi} with $\mathcal{Q}$, in the present setting it holds that
\begin{equation}
\mathcal{Q}(\mathbf{Y}) = T_{\text{max}}(\mathbf{Y}) = f\left(\mathbf{T}(\mathbf{Y})\right). 
\end{equation}

\section{High Dimensional Uncertainty Quantification}
We are interested in computing statistical moments of a \gls{qoi} $\mathcal{Q}(\mathbf{Y})$, such as its expectation value $\mathbb{E}[\mathcal{Q}] = \mu_{\mathcal{Q}}$ and its standard deviation $\sigma_{\mathcal{Q}}$.
A common way to compute these is with sampling methods, e.g. the \glsfirst{mc} method, which however converges slowly.
While more efficient methods are available, they suffer from the so-called \emph{curse of dimensionality}, i.e. their computational work increases, often exponentially, with respect to the number of input \glspl{rv}.
We are therefore interested in methods that either alleviate or break the curse, while maintaining good convergence properties.
Moreover, we are interested in \emph{non-intrusive} methods, i.e. methods that allow the use of deterministic numerical methods for the solution of the semi-discrete problem, i.e. after its discretization in space and time, without any modifications. 
Two methods have been found to fulfill these criteria and are presented in the following sections, namely the \emph{\glsfirst{stc}} on \emph{sparse grids (SG)}  and the \emph{\glsfirst{tt} decomposition}.

\subsection{Stochastic Collocation on Sparse Grids} \label{subsec:SC-SG}
% Stochastic collocation on tensor grids
In the \gls{stc} method, introduced in \cite{BabSC,XiuHigh}, the deterministic problem is solved exactly on a set of \gls{rv} realizations which are called \emph{collocation points}.
The choice of the collocation points $\mathbf{y}^{(i)}$ is based on the distributions of the input \glspl{rv}. 
We seek to construct a global polynomial approximation of the \gls{qoi} $\mathcal{Q}$, such that
\begin{equation}
\label{eq:approx}
\mathcal{Q}(\mathbf{y}) \approx \tilde{\mathcal{Q}}(\mathbf{y}) = \sum_{m=1}^M \alpha_m L_m(\mathbf{y}),
\end{equation}
for suitable multivariate polynomials $\{L_m(\mathbf{y})\}_{m=1}^M$ and coefficients $\alpha_m$. A brief summary of the method follows, for more extensive presentations see \cite{BabSC, MotSC}.

First, we introduce a tensor-product space $\mathcal{P}_\ell$, which is the span of tensor-product polynomials with degree at most $\mathbf{p} = \left[p_1, \dots, p_N\right]$.
The positive integer $\ell$ is called the \emph{approximation level}.
It holds that $\mathcal{P}_\ell = \bigotimes_{n=1}^N \mathcal{P}_{p_n}$, where $\mathcal{P}_{p_n}$ is the span of univariate polynomials up to degree $p_n$. 
We denote with $y_n^{(i_n)}$, $0 \leq i_n \leq p_n$, the $(p_n+1)$ roots of the polynomials of degree $\left(p_n+1\right)$ living in $\mathcal{P}_{p_n}$, which are orthogonal with respect to the univariate PDF $\rho_n(y_n)$.

Next, let a multi-index $\mathbf{j} = [j_1, \dots, j_N]$, such that the polynomial degrees $p_n$ are given as functions of the indices $j_n$, i.e. $p_n = p(j_n)$.
The multi-dimensional Lagrange interpolation operator corresponding to the multi-index $\mathbf{j}$ reads
\begin{equation}
\mathcal{I}_{\mathbf{j},N} \mathcal{Q}\left(\mathbf{y}\right) = \sum_{i_1=0}^{p(j_1)} \dots \sum_{i_N=0}^{p(j_N)} \mathcal{Q}\left(y_{1}^{(i_1)},\dots,y_{N}^{(i_N)} \right)  \prod_{n=1}^N l_{n,i_n} \left(y_n \right),
\end{equation}
where $\{l_{n,i_n}(y_n)\}_{i_n=0}^{p(j_n)}$ are univariate Lagrange polynomials of degree $p(j_n)$, such that
\begin{equation}
l_{n,i_n}(y_n) = \prod_{k=0, k \neq i_n}^{p(j_n)}\frac{y_n - y_n^{(k)}}{y_n^{(i_n)} - y_n^{(k)}}.
\end{equation}

For $p(j_n) = j_n$ and $\mathbf{j} = \left[ \ell, \ell, \dots, \ell \right]$, we obtain the \emph{isotropic full tensor-product space}, i.e. in each direction we take all polynomials of degree at most $p(\ell) = \ell$. 
In this case, the dimension of $\mathcal{P}_{\ell}$ is $\dim\left[\mathcal{P}_\ell \right] = \left(\ell +1\right)^N$.
Accordingly, the collocation points $\mathbf{y}^{(i)} = \left(y_1^{(i_1)}, y_2^{(i_2)}, \dots, y_N^{(i_N)}\right)$ are given as the tensor grid
\begin{equation}
H_{\mathbf{j},N}^{\mathrm{TP}} := \{ y_1^{(0)},\dots,y_1^{(\ell)} \} \times \{ y_2^{(0)},\dots,y_2^{(\ell)} \} \times \dots \times \{y_N^{(0)},\dots,y_N^{(\ell)} \},
\end{equation}
and the corresponding Lagrange interpolation operator is denoted with $\mathcal{I}_{\ell, N}$.
The global index $i$ is associated with the local indices $i_n$ as in \cite{MotSC}.

For the computation of statistical moments, we use a multi-dimensional Gauss quadrature formula for integral approximation.
The number of quadrature points per dimension is equal to $\left(\ell+1\right)$.
Then, taking as collocation points the tensor grid created by the one-dimensional quadrature abscissas, we write
\begin{equation}
\mathbb{E}[\mathcal{Q}] = %\approx 
%\mathbb{E}[\tilde{\mathcal{Q}}] =
\int_{\Omega} \mathcal{Q}(\mathbf{y}) \rho(\mathbf{y}) d\mathbf{y} \approx 
\sum_{i=1}^I w_i \mathcal{Q}(\mathbf{y}^{(i)}),
\label{eq:GaussQuad}
\end{equation}
where $\Omega$ denotes the set of random outcomes and the weights $w_i$ are given as
\begin{equation*}
w_i = \prod_{n=1}^N w_{i_n} \:\:\:\:\: \text{and} \:\:\:\:\: w_{i_n} = \int_{\Omega_n} l_{n,i_n}^2(y_n) \rho_n(y_n) dy_n.
\end{equation*}

% Sparse grids part
However, the tensor grid approach soon becomes intractable, as $\dim\left[\mathcal{P}_\ell \right]$ and the number of collocation points increase exponentially with the number of \glspl{rv}. 
Therefore, collocation in high dimensions exploits \emph{sparse grids} \cite{BarSG, BungSparse, NobSparse}.
For simplicity, only isotropic \emph{total degree} and \emph{Smolyak} grids are presented.
Readers interested in anisotropic grids are referred to \cite{NobAniso} and the references therein.
For the total degree case, we set $p(j) = j$, while for the Smolyak grid case we set $p(j) = 2^j$ for $j>0$, and $p(0)=0$.
For both cases we require that $|\mathbf{j}| \leq p(\ell)$, where $|\mathbf{j}| = \sum_{n=1}^N j_n$.
Then, the sparse interpolation is obtained by the isotropic Smolyak formula
\begin{equation}
\mathcal{A}_{\ell,N} \mathcal{Q}\left(\mathbf{y}\right) = \sum_{\ell-N+1 \leq |\mathbf{j}| \leq \ell} (-1)^{\ell-|\mathbf{j}|} \binom{N-1}{\ell - |\mathbf{j}|} \mathcal{I}_{\mathbf{j},N} \mathcal{Q} \left(\mathbf{y} \right).
\label{eq:int_sparse}
\end{equation}
The associated sparse grids are denoted with $H_{\ell, N}^{\text{TD}}$ and $H_{\ell, N}^{\text{SM}}$, for the total degree and the Smolyak case, respectively.

\subsection{Tensor Train Decomposition} \label{subsec:TTdec}
Considering again the full tensor-product space introduced in Section \ref{subsec:SC-SG} and the quadrature formula (\ref{eq:GaussQuad}), we define the $N$-dimensional arrays, or $N$-\emph{tensors}, $\mathbfcal{Q}$, $\mathbfcal{W}$ $\in \mathbb{R}^{I_1 \times \cdots \times I_N}$, with elements
\begin{align}
\mathbfcal{Q}\left(i_1, i_2, \dots, i_N\right) := \mathcal{Q}\left(y_1^{(i_1)}, y_2^{(i_2)}, \cdots, y_N^{(i_N)}\right) 
\:\: \text{and} \:\:
\mathbfcal{W}\left(i_1, i_2, \dots, i_N\right) := w_{i_1} w_{i_2} \cdots w_{i_N},
\label{eq:tensors}
\end{align}
where $i_n = 1, \dots, I_n$.
In the isotropic case, it holds that $I_1 = \cdots = I_N = \left(\ell +1\right)$.

Then, (\ref{eq:GaussQuad}) can be re-written in the form of the inner product between $\mathbfcal{Q}$ and $\mathbfcal{W}$, i.e.
\begin{align}
\mathbb{E}[\mathcal{Q}] \approx 
\mathbb{E}[\tilde{\mathcal{Q}}] =
\langle \mathbfcal{Q}, \mathbfcal{W} \rangle := \sum_{i_1, \dots, i_N} \mathbfcal{Q}\left(i_1, \dots, i_N\right) \mathbfcal{W}\left(i_1, \dots, i_N\right).
\label{eq:tensorQuad}
\end{align}
For the comfort of the reader, the standard notation for tensor and matrix elements is substituted by a MATLAB-like notation, such that $\mathbfcal{Q}(i_1, \dots, i_N) = q_{i_1, \dots, i_N}$ and $\mathbfcal{W}(i_1, \dots, i_N) = w_{i_1, \dots, i_N}$. 
Obviously, the storage needs of $\mathbfcal{Q}$ and $\mathbfcal{W}$ and the computational work involving these tensors increases exponentially with respect to $N$.
However, this curse of dimensionality can be overcome using \emph{low-rank tensor decomposition} methods, which yield representations or approximations of tensors, in formats requiring much less storage and allowing efficient computations.

Low-rank decomposition is a problem which is well-studied for the 2-dimensional, i.e. matrix, case.
Using element-wise notation, a rank-$R$ decomposition of a matrix $\mathbf{A} \in \mathbb{R}^{I_1 \times I_2}$ reads  
\begin{align}
\mathbf{A}(i_1, i_2) \approx \tilde{\mathbf{A}}(i_1, i_2) = \sum_{r=1}^R \mathbf{U}(i_1, r) \mathbf{V}(r, i_2).
\label{eq:low-rank decomp}
\end{align}
If $R < \text{rank}\left(\mathbf{A}\right)$, the decomposition yields a rank-$R$ approximation of the matrix. 
The best rank-$R$ approximation in the Frobenius-norm sense can be obtained by the truncated \emph{\gls{svd}}, such that $\tilde{\mathbf{A}} = \mathbf{U}_R \mathbf{S}_R \mathbf{V}_R^{*}$, which however is expensive to compute.
A computationally less expensive alternative to the \gls{svd} is the \emph{skeleton} or \emph{cross} approximation \cite{BebACA, GorPseudo}, defined as
\begin{equation}
\mathbf{A} \approx \tilde{\mathbf{A}} = \mathbf{A}\left(:, \mathcal{I}^2\right) \mathbf{A}\left(\mathcal{I}^1, \mathcal{I}^2\right)^{-1} \mathbf{A}\left(\mathcal{I}^1, : \right),
\end{equation}
or equivalently, in element-wise notation 
\begin{align}
\mathbf{A}(i_1, i_2) \approx \tilde{\mathbf{A}}(i_1, i_2) = \sum_{s,t = 1}^R \mathbf{A}\left(i_1, i_t^2\right) \mathbf{A}\left(i_s^1, i_t^2\right)^{-1} \mathbf{A}\left(i_s^1, i_2\right),
\end{align}
where $\mathcal{I}^1 = \{ i_s^1 \}_{s=1}^R$ and $\mathcal{I}^2 = \{ i_t^2 \}_{t=1}^R$ denote row and column index sets, respectively.
The approximation matrix $\tilde{\mathbf{A}}$ can be computed by touching only $(I_1+I_2)R - R^2$ entries of the original matrix. 
Therefore, this approach is particularly efficient for function-related matrices, where $\mathbf{A}(i_1, i_2) := f(x_{i_1}, x_{i_2})$, with $x: i_j \in \mathbb{N} \mapsto x_{i_j} \in \mathbb{R}$ and assuming that $f(x_1, x_2)$ is a smooth function.
The accuracy of the approximation depends on the choice of index sets $\mathcal{I}^1, \mathcal{I}^2$. 
For the best possible approximation, $\mathcal{I}^1$ and $\mathcal{I}^2$ must be chosen such that $\mathbf{A}\left(\mathcal{I}^1, \mathcal{I}^2\right)$ is a submatrix of \emph{maximal volume}, meaning that it has the maximal in modulus determinant over all $R \times R$ submatrices of $\mathbf{A}$.
This task however is computationally challenging and often unfeasible. 
In practice only \emph{quasi}-maximal volume submatrices \cite{GorHow} are required.
The choice of index sets such that the maximality or quasi-maximality criterion is satisfied, is often denoted with $\left[\mathcal{I}^1, \mathcal{I}^2\right] = \mathbf{maxvol}\left[\mathbf{A}\right]$.

Various efforts have been made to generalize the low-rank decomposition (\ref{eq:low-rank decomp}) from matrices to tensors \cite{GrasRev, HackBook, KhorRev, KoldaRev}.
Here, we consider only the \emph{\glsfirst{tt}} decomposition \cite{OseTTdec}, which for an $N$-tensor $\mathbfcal{A} \in \mathbb{R}^{I_1 \times \cdots \times I_N}$ is defined as
\begin{align}
\mathbfcal{A}(i_1, \dots, i_N) \approx \mathbfcal{A}_{\text{TT}}(i_1, \dots, i_N)  
&= \sum_{r_0,\dots, r_N}^{R_0,\dots,R_N} \mathbfcal{G}_1(r_0, i_1, r_1) \mathbfcal{G}_2(r_1, i_2, r_2) \cdots \mathbfcal{G}_N(r_{N-1}, i_N, r_N) \nonumber \\
&= \sum_{\mathbf{r}}^{\mathbf{R}} \prod_{n=1}^N \mathbfcal{G}_n\left(r_{n-1}, i_n, r_n\right),
\label{eq:TTdecomp}
\end{align}
where $\mathbfcal{G}_n \in \mathbb{R}^{R_{n-1} \times I_n \times R_n}$ are the \emph{\gls{tt}-cores}, $\mathbf{R} = (R_0, R_1, \dots, R_N)$ are the \emph{\gls{tt}-ranks}, $i_n = 1,\dots,I_n$ are the \emph{mode indices} and $r_n = 1,\dots,R_n$ are auxiliary indices.
It holds that $R_0 = R_N = 1$.
The \gls{tt} decomposition is also known as the \emph{Matrix Product State (MPS)} format, due to fact that it can be written as
\begin{align}
%\mathbfcal{A}(i_1, \dots, i_d) \approx 
\mathbfcal{A}_{\text{TT}}(i_1, \dots, i_N) = \mathbf{G}_1(i_1) \mathbf{G}_2(i_2) \cdots \mathbf{G}_N(i_N),
\end{align}
where $\mathbf{G}_n(i_n)$ are $R_{n-1} \times R_n$ matrices, one for each index $i_n$.

We define the multi-indices $i_{\leq n} := \left(i_1, \dots, i_n\right)$ and $i_{>n} := \left(i_{n+1},\dots,i_N\right)$, and the $n$-th \emph{unfolding} or \emph{matricization} as the matrix $\mathbf{A}_n \in \mathbb{R}^{\left(I_1 \cdots I_n\right) \times \left(I_{n+1} \cdots I_N\right)}$ with elements
\begin{align}
\mathbf{A}_n\left( i_{\leq n}, i_{>n}\right) = \mathbf{A}_n\left(\overline{i_1, \dots, i_n}, \overline{i_{n+1}, \dots, i_N}\right) = \mathbfcal{A}(i_1, \dots, i_N).
\end{align}
The \gls{tt} decomposition is exact for $R_n = \text{rank}\left(\mathbf{A}_n\right)$, and approximate for $R_n < \text{rank}\left(\mathbf{A}_n\right)$. 
Setting $\max(I_n) = I$ and $\max(R_n) = R$, $n=1,\dots,N$, the storage of a tensor in the \gls{tt} format requires $\mathcal{O}\left( NIR^2\right)$ elements, which is substantially less than the $\mathcal{O}\left(I^N\right)$ entries of the full tensor and especially helpful in the case of function-related tensors.
Storage and computational operations in the \gls{tt} format scale \emph{linearly} with the number of dimensions $N$, thus breaking the curse of dimensionality \cite{OseTTdec}.

The \gls{tt} approximation is typically computed by solving the minimization problem 
\begin{align}
\min_{\mathbfcal{G}_1,\dots,\mathbfcal{G}_N} \|\mathbfcal{A} - \mathbfcal{A}_{\text{TT}}\|_F,
\label{eq:TToptim}
\end{align}
where $\| \cdot \|_{F}$ denotes the Frobenius norm.
In \cite{OseTTcross}, an approach based on the \emph{alternating least squares (ALS)} algorithm is proposed to solve (\ref{eq:TToptim}).
The idea relies on organizing the ALS iteration in \emph{left-to-right} and \emph{right-to-left} sweeps and fixing at each step all TT-cores but one, thus reducing the number of unknowns to $R_{n-1}\times I_n \times R_n$ at a time.
In this, so-called, \emph{TT-cross} algorithm, cross approximation replaces least-squares in optimizing the TT-core at each step.
We denote with $\mathcal{I}^{\leq n} = \{i_{s_n}^{\leq n}\}_{s_n=1}^{R_n}$ and $\mathcal{I}^{> n} = \{i_{t_n}^{> n}\}_{t_n=1}^{R_n}$ the index sets with the positions of the $R_n$ rows and columns of matricization $\mathbf{A}_n$, such that $\left[ \mathbfcal{A}\left(i_{s_n}^{\leq n}, i_{t_n}^{>n}\right) \right]_{s_n, t_n = 1}^{R_n} = \left[\mathbfcal{A}\left(\mathcal{I}^{\leq n}, \mathcal{I}^{>n}\right)\right]$. 
Further assuming that  $\left[\mathcal{I}^{\leq n}, \mathcal{I}^{>n}\right] = \mathbf{maxvol}\left[\mathbf{A}_n\right]$, the generalized TT-cross approximation from \cite{OseTTcross} reads 
\begin{align}
\mathbfcal{A}_{\text{TT}}(i_1, \dots, i_N) %&= 
%\sum_{\mathbf{s}, \mathbf{t}} 
%\mathbfcal{A}\left(i_1, \mathcal{I}_{t_1}^{>1}\right) 
%\left[\mathbfcal{A}\left(\mathcal{I}_{s_1}^{\leq 1}, \mathcal{I}_{t_1}^{>1}\right) \right]^{-1}
%\mathbfcal{A}\left(\mathcal{I}_{s_1}^{\leq 1}, i_2, \mathcal{I}_{t_2}^{>2}\right) 
%\cdots
%\mathbfcal{A}\left(\mathcal{I}_{s_{d-1}}^{\leq d-1}, i_d\right) \nonumber \\
&= \sum_{\mathbf{s}, \mathbf{t}} \prod_{n=1}^N 
\mathbfcal{A}\left(i_{s_{n-1}}^{\leq k-1}, i_n, i_{t_n}^{>n}\right)
\left[\mathbfcal{A}\left(i_{s_n}^{\leq n}, i_{t_n}^{>n}\right) \right]^{-1}.
\end{align}
Border sets $\mathcal{I}^{\leq 0} = \emptyset$ and $\mathcal{I}^{>N} = \emptyset$ have been introduced to unify the notation.
The total complexity of the TT-cross algorithm is $\mathcal{O}\left( NIR^3 \right)$, i.e. scales linearly with respect to the number of dimensions $N$.

TT-cross has the drawback that the \gls{tt}-ranks $R_n$ have to be guessed in advance.
Underestimations of the \gls{tt}-ranks may lead to poor approximations, while overestimations increase computational costs.
A more efficient and \emph{rank-revealing} method is presented in \cite{SavFast}.
The corresponding algorithm is based on the \emph{Density Matrix Renormalization Group (DMRG)} numerical technique from quantum chemistry, which can be seen as an idea of constrained optimization in the MPS format.
In this \emph{\gls{tt}-RC (Renormalization Cross)} algorithm, optimization is performed over two \gls{tt}-cores, $\mathbf{G}_n$ and $\mathbf{G}_{n+1}$, simultaneously.
At step $n$ of the sweeps, the supercore $\mathbf{W}_n\left(i_n, i_{n+1}\right) = \mathbf{G}_n\left(i_n\right) \mathbf{G}_{n+1}\left(i_{n+1}\right)$ is found by solving (\ref{eq:TToptim}).
Matrix $\mathbf{W}_n$ is of size $R_{n-1} I_n \times I_{n+1} R_{n+1}$, hence it does not include any information regarding rank $R_n$.
Matrices $\mathbf{G}_n$ and $\mathbf{G}_{n+1}$ are recovered using a truncated-\gls{svd} of $\mathbf{W}_n$. 
This, so-called, \emph{decimation} step provides us with the \gls{tt}-rank $R_n$, thus the rank-revealing property of the method.

In \cite{SavQuasi}, a greedy approach to \gls{tt}-cross approximation is discussed, similar to the \emph{adaptive cross approximation (ACA)} \cite{BebACA} method for matrices.
The proposed algorithm, called \emph{greedy-TT-cross}, modifies the TT-RC algorithm and adds at each step a new pivot index, based on where the error of the current approximation is quasi-maximum in the current step's supercore $\mathbf{W}_n$.
Its total computational complexity is also $\mathcal{O}(NIR^3)$, i.e. it remains linear with respect to the number of dimensions $N$.

\section{Numerical Results}
Simulations of the electrothermal field problem presented in Section \ref{sec:ETProblem} were performed.
The implicit Euler method was used for time integration.
In total, the initial 3.5 seconds of the \gls{ic}'s duty cycle were simulated, divided into 51 equidistant time-steps.
The total amount of degrees of freedom due to the \gls{fit} discretization was 10516, for each time-step.

First, a \gls{mc} simulation with 40000 realizations of the input \glspl{rv} was performed.
Each realization corresponds to a single call of the deterministic solver which solves (\ref{eq:ext1}) and (\ref{eq:ext2}).
%The corresponding convergence plots regarding the mean value $\mathbb{E}\left[T_{\text{max}}\right] = \mu_{T_{\text{max}}}$ and the standard deviation $\sigma_{T_{\text{max}}}$ of the maximum bondwire temperature are presented in Figure \ref{fig:MCres}. 
For both $\mathbb{E}\left[T_{\text{max}}\right]$ and $\sigma_{T_{\text{max}}}$ we have convergence up to the second digit.
The most accurate values of the mean and the standard deviation of $T_{\text{max}}$, i.e. those computed with the full set of 40000 samples, are $\mathbb{E}\left[T_{\text{max}}\right]^{\text{MC}} = \mu_{T_\text{max}}^{\text{MC}} = 544.42 \: \text{K}$ and $\sigma_{T_{\text{max}}}^{\text{MC}} = 4.07 \: \text{K}$, respectively.
The main question is whether we can reach the most accurate \gls{mc} mean and standard deviation values with significantly less calls to the solver, by applying either the \gls{stc}-\gls{sg} or the \gls{tt} decomposition.
\begin{comment}
\begin{figure}[h]
\begin{subfigure}[b]{0.5\textwidth}
\centering
\includegraphics[width=0.9\textwidth]{mean_Tmax_10.pdf}
\caption{Mean $T_{\text{max}}$.}   
\label{fig:meanTmax}
\end{subfigure}
%%%
\begin{subfigure}[b]{0.5\textwidth}
\centering
\includegraphics[width=0.9\textwidth]{std_Tmax_10.pdf}
\caption{Standard deviation of $T_{\text{max}}$.}
\label{fig:stdTmax}
\end{subfigure}
\caption{Monte Carlo convergence plots.} 
\label{fig:MCres}
\end{figure}
\end{comment}

The \gls{stc}-\gls{sg} method was implemented using the \emph{Dakota toolkit} \cite{Dakota}.
Approximation levels $\ell=1,2,3$ have been considered for the case of Smolyak sparse grids.
The quadrature abscissas chosen for (\ref{eq:GaussQuad}) are the \emph{Clenshaw-Curtis (CC)}.
The number of calls to the deterministic solver is equal to the number of collocation points.
The relative errors for the mean value and the standard deviation are computed with $\mu_{T_\text{max}}^{\text{MC}}$ and $\sigma_{T_{\text{max}}}^{\text{MC}}$ as the reference values, respectively.
The corresponding data and results for all cases are presented in Table \ref{tab:sc-sg}.
\begin{comment}
% Nobile's Sparse grid toolkit results
\begin{table}[h!]  \centering        
\begin{tabular}{|c|cc|cc|cc|c|}
\hline
\multicolumn {1}{|c|}{$\ell$} & \multicolumn {2}{c|}{Colloc. points} & \multicolumn {2}{c|}{$\epsilon_{\mu, \text{SG}} \: \left(\%\right)$} & \multicolumn {2}{c|}{$\epsilon_{\sigma, \text{SG}} \: \left(\%\right)$} \bigstrut[t] \\
\hline
-- & SM (CC) & TD (GL) & SM (CC) & TD (GL) & SM (CC) & TD (GL) \\ 
\hline
1 & 25    & 25       & $<$ 1.0 & $<$ 1.0 & 6.71  & 11.34 \bigstrut[t] \\
2 & 313   & 313      & $<$ 1.0 & $<$ 1.0 & 10.34 & 4.64  \\
3 & 2649  & 2649     & $<$ 1.0 & $<$ 1.0 & 16.69 & 4.43  \\
4 & 17265 & 17217    & $<$ 1.0 & $<$ 1.0 & 20.38 & 17.37 \bigstrut[b]\\
\hline
\end{tabular}
\caption{Stochastic collocation on sparse grids results.}
\label{tab:sc-sg}
\end{table}
\end{comment}
\begin{table}[h!]  \centering        
    \begin{tabular}{|c|c|c|}
        \hline
        \multicolumn {1}{|c|}{$\ell$} & \multicolumn {1}{c|}{Collocation points} & \multicolumn {1}{c|}{St. dev. rel. error $\epsilon_{\sigma, \text{SG}} \: \left(\%\right)$} \bigstrut[t] \\
        \hline
        1 & 25   & 3.50  \bigstrut[t] \\
        2 & 337  & 3.19  \\
        3 & 3249 & 3.36  \bigstrut[b] \\
        %4 & 17265 & 17217    & $<$ 1.0 & $<$ 1.0 & 20.38 & 17.37 \bigstrut[b]\\
        \hline
    \end{tabular}
    \caption{Results for the stochastic collocation on sparse grids.}
    \label{tab:sc-sg}
\end{table}

The relative errors for the mean value are consistently below $1\%$ and therefore omitted.
The relative errors for the standard deviation  are consistently between $3\%-4\%$, even for a the case of $\ell=1$, i.e. with only 25 collocation points.
Therefore, a good approximation of the reference mean and standard deviation values can be obtained using stochastic collocation on sparse grids, with significantly less solver calls.
However, some oscillation in the relative error of the standard deviation still remains, therefore it can be deduced that more approximation levels are required until full convergence is reached.

For the implementation of the \gls{tt} decomposition approach, the MATLAB \emph{TT-Toolbox} \cite{TTtoolbox} was used.
We denote with $\otimes$ the outer product and with $\mathbf{w}_n \in \mathbb{R}^{(\ell+1)}$ the vectors containing all quadrature weights for the corresponding \glspl{rv} $Y_n$, such that $\mathbf{w}_n\left(i_n\right) = w_{i_n}$. 
Then, the $N$-tensor $\mathbfcal{W}$ from (\ref{eq:tensors}) can be written as
\begin{equation}
\mathbfcal{W} = \mathbf{w}_1 \otimes \cdots \otimes \mathbf{w}_N,
\label{eq:Wrank1}
\end{equation}
or equivalently, in element-wise notation
\begin{equation}
\mathbfcal{W}\left(i_1, \dots, i_N\right) = \mathbf{w}_1(i_1) \cdots \mathbf{w}_N(i_N). 
\label{eq:Wrank2}
\end{equation}
The formats in (\ref{eq:Wrank1}), (\ref{eq:Wrank2}) are equivalent to an exact \gls{tt} decomposition, i.e. $\mathbfcal{W}_{\text{TT}} = \mathbfcal{W}$, with all \gls{tt}-ranks equal to 1.
Therefore, storing $\mathbfcal{W}_{\text{TT}}$ requires only $N\left(\ell+1\right)$ elements.

Since $\mathbfcal{W}_{\text{TT}}$ is an exact \gls{tt} decomposition, the computation of statistical moments relies predominantly on the approximation of the $N$-tensor $\mathbfcal{Q}$, which is obtained using the \emph{greedy-TT-cross} algorithm.
The algorithm is run for an increasing number of allowed sweeps, which in turn leads to increasing TT-ranks, larger interpolation index sets and thus better approximations, at the cost of more calls to the deterministic solver.
Each approximation $\mathbfcal{Q}_{\text{TT}}$ yields different results for $\mathbb{E}[T_{\text{max}}] = \mu_{T_{\text{max}}}$ and $\sigma_{T_{\text{max}}}$. 
The respective results are presented in Table \ref{tab:ttcross123}.
\begin{table}[h!]  \centering        
    \begin{tabular}{|c|ccc|c|ccc|}
        \hline
        \multicolumn {1}{|c|}{Sweeps} & \multicolumn {3}{c|}{Solver calls} & \multicolumn {1}{c|}{Max. TT-rank $R_{n,\text{max}}$} & \multicolumn {3}{c|}{St. dev. rel. error $\epsilon_{\sigma, \text{TT}} \: \left(\%\right)$}  
        \bigstrut[t] \\
        \hline
        -- & $\ell=1$ & $\ell=2$ & $\ell=3$ & -- & $\ell=1$ & $\ell=2$ & $\ell=3$ \\
        \hline
        1  & 154  & 242  & 327  & 2  & 1.96    & 12.25   & 18.86 \bigstrut[t] \\
        2  & 291  & 491  & 670  & 3  & 6.52    & 2.56    & 14.03 \\
        3  & 460  & 772  & 1133 & 4  & 6.90    & 125.43  & 13.63 \\
        4  & 647  & 1130 & 1616 & 5  & 56.42   & 11.62   & 10.66 \\
        5  & 846  & 1529 & 2200 & 6  & 3.05    & 1.49    & 12.49 \\ 
        6  & 1018 & 1938 & 2980 & 7  & 1.05    & 1.49    & 5.30  \\
        7  & 1264 & 2790 & 3713 & 8  & $<$ 1.0 & 2.70    & 4.25  \\
        8  & 1497 & 3284 & 4536 & 9  & $<$ 1.0 & $<$ 1.0 & 2.26  \\ 
        9  & 1753 & 3981 & 5833 & 10 & 1.10    & 2.20    & 3.19  \\
        10 & 2028 & 4550 & 6875 & 11 & $<$ 1.0 & 1.48    & 1.90  \bigstrut[b]\\
        \hline
    \end{tabular}
    \caption{Results for the tensor train decomposition.}
    \label{tab:ttcross123}
\end{table}

As can be observed, the maximum TT-rank increases with the number of allowed sweeps.
The maximum TT-rank is an indicator of the storage needs for $\mathbfcal{Q}$ in the TT format.
As discussed in Section \ref{subsec:TTdec}, $\mathcal{O}\left(N\left(\ell+1\right)R_{n, \text{max}}^2\right)$ elements are required to store a tensor in the TT format.
Here, storing $\mathbfcal{Q}_{\text{TT}}$ requires much less space than this limit, because the maximum \gls{tt}-rank appears only in few cores.

As in the sparse grids case, the mean value relative errors are consistently below $1\%$ for all considered approximation levels and allowed sweeps, and are therefore omitted. 
Observing the standard deviation relative errors, it seems that accurate approximations of tensor $\mathbfcal{Q}$ are obtained when the allowed sweeps are more than 5 for $\ell=1,2$ and more than 7 for $\ell=3$. 
After that point, deviations from the reference value are consistently less than $5\%$. 
Hence, after a certain number of sweeps, the algorithm has located interpolation index sets which provide a good \gls{tt} approximation $\mathbfcal{Q}_{\text{TT}} \approx \mathbfcal{Q}$, thus leading to accurate mean and standard deviation values.
It can be observed that accuracies superior to those of the \gls{stc}-\gls{sg} can be obtained, albeit at the cost of more solver calls.   

It can be observed that, in the \gls{tt}-based approach, the most accurate results are obtained for $\ell=1$, which indicates low-level interactions between the \glspl{rv}.
This should be investigated further, since it could be the case that high approximation levels do not really contribute to better accuracies.
A possible question is whether the use of full tensor-product spaces of lower order would yield better approximation results.
With respect to the more accurate results of the TT decomposition approach, it should be noted that isotropic sparse grids are \emph{a priori} constructed for a \emph{class} of smooth-enough functions, while the \gls{tt} approximation is built for a specific function.
This is a structural disadvantage for the \gls{stc}-\gls{sg} method, which should be taken into consideration.
Adaptive construction of sparse grids, i.e. grids tailored to a specific function, is a possible extension of this work. 

\section{Conclusion}
Uncertainty quantification methods based on stochastic collocation on sparse grids and on tensor train decompositions were implemented for an electrothermal problem with 12 random variables.
Accurate results for the first two statistical moments, i.e. mean and standard deviation, of the considered quantity of interest were obtained with both methods, with significantly less solver calls than using the Monte Carlo method. 
Stochastic collocation on sparse grids appears to be the computationally most efficient method for the considered problem setting.
At the cost of higher computational burden, the tensor train approach yields results of superior accuracy, compared to stochastic collocation.
However, the aforementioned structural advantage of the tensor train decomposition has to be taken into consideration.
Extensions of the present work shall handle problems involving an increased number of random variables, with the goal of determining the efficiency of both approaches, especially of the more novel tensor train decomposition, when the impact of the curse of dimensionality is more pronounced.

\section*{Acknowledgement}
This work was supported by the Graduate School of Computational Engineering, Technische Universit\"at Darmstadt. 
Part of this work was supported by the fp7-nanoCOPS (Nanoelectronic Coupled Problems Solutions) project.

\end{document}